\documentclass{article}
\usepackage{spconf,amsmath,graphicx}
\usepackage[table,xcdraw]{xcolor}
\usepackage{multirow}
\usepackage{cite}
\title{Low-resource expressive text-to-speech using data augmentation\vspace{-1mm}}
\name{Goeric Huybrechts, Thomas Merritt, Giulia Comini, Bartek Perz, Raahil Shah, Jaime Lorenzo-Trueba\vspace{-1mm}}
\address{Amazon Alexa - TTS Research, Cambridge, United Kingdom\\
\{huybrech, thommer, gcomini, perzbart, raahshah, truebaj\}@amazon.com\vspace{-2mm}}
\begin{document}
%
\maketitle
\begin{abstract}
While recent neural text-to-speech (TTS) systems perform remarkably well, they typically require a substantial amount of recordings from the target speaker reading in the desired speaking style. In this work, we present a novel 3-step methodology to circumvent the costly operation of recording large amounts of target data in order to build expressive style voices with as little as 15 minutes of such recordings. First, we augment data via voice conversion by leveraging recordings in the desired speaking style from other speakers. Next, we use that synthetic data on top of the available recordings to train a TTS model. Finally, we fine-tune that model to further increase quality. Our evaluations show that the proposed changes bring significant improvements over non-augmented models across many perceived aspects of synthesised speech. We demonstrate the proposed approach on 2 styles (newscaster and conversational), on various speakers, and on both single and multi-speaker models, illustrating the robustness of our approach.\footnote{\tiny{Samples available on www.amazon.science/low-resource-expressive-text-to-speech-using-data-augmentation}}
\end{abstract}
\vspace*{-4mm}
\section{Introduction}
\label{sec:intro}
\vspace*{-3mm}

Recently, the naturalness and quality of speech generated with neural text-to-speech (TTS) systems has improved significantly \cite{oord2016wavenet, wang2017tacotron, sotelo2017char2wav, skerry2018towards, kalchbrenner2018efficient, oord2018parallel}. However, as concluded in \cite{chung2019semi}, high-quality state-of-the-art models such as Tacotron \cite{wang2017tacotron} require at least 10 hours of text and speech pairs. Since collecting data is a costly operation, the need for alternatives is high.

Research that focuses on low-resource TTS tries to mitigate the effects of limited data via multi-speaker modelling and transfer learning \cite{gibiansky2017deep, jia2018transfer, tits2019exploring, latorre2019effect, chen2019end, chung2019semi, zhang2020unsupervised}. By transferring knowledge gained from high-resource speakers, the quality of low-resource systems improves. This is not the only way to tackle the problem of scarce data though.

Data augmentation has been used across fields in machine learning to overcome the aforementioned issue. Especially in computer vision, creating synthetic data to train better models has been thoroughly investigated \cite{perez2017effectiveness, mikolajczyk2018data, shorten2019survey}. Data augmentation has also been researched in the speech field, particularly in the field of automatic speech recognition \cite{ko2015audio, ko2017study, park2019specaugment}. However, despite the large amount of research conducted to create synthetic speech data, only \cite{xu2020lrspeech} seems to leverage it for non-expressive TTS end goals. A multi-speaker TTS model is trained, which is used to create synthetic data for the target speaker, which in turn is used to train a new dedicated single-speaker TTS model. This assumes the low-resource multi-speaker model can synthesise speech for the target speaker (with natural prosody and good signal quality), a non-trivial assumption.

A more effective way to augment data in the speech field which bypasses this constraint can be achieved through voice conversion (VC). VC aims to convert an utterance spoken by a source speaker as though it was read by a different target speaker, while maintaining all other features such as linguistic content and prosody. Much research has already been conducted on VC \cite{mohammadi2017overview, hsu2017voice, lorenzo2018voice, kaneko2019cyclegan}. In this paper, we chose to use Copycat \cite{karlapati2020copycat}, known for its robustness and excellent VC quality. Nevertheless, we expect results to be applicable to other high-quality VC techniques.
 
In this work, we present a novel methodology which results in high-quality TTS models, despite using only a small amount of expressive samples for the target speaker. We propose a 3-step solution involving voice-converted data and fine-tuning. The voice-converted data resembles the target recordings more than any other auxiliary data, regardless of whether this data belongs to another style or speaker. Hence, the model is trained on much more data coming from a similar distribution and can more reliably produce the desired type of speech. The fine-tuning, in turn, allows the model to focus on the actual target space more closely. Our methodology shows significant improvements across many use cases, illustrating its robustness. To the best of our knowledge, our proposed methodology is the first one that uses a VC model to augment data with the aim to train better low-resource TTS models.
 
\vspace*{-2mm}
\section{Methodology}
\label{sec:methodology}
\vspace*{-2mm}

The proposed methodology consists of 3 steps. In the first step, we create synthetic data by leveraging a VC model \cite{karlapati2020copycat}. We augment recordings from supporting speakers recorded in the target speaking style to the identity of our target speaker. In the second step, the newly-generated synthetic data is used in addition to the target speaker recordings to train a state-of-the-art TTS model. Finally, fine-tuning of the TTS model is performed to achieve higher quality. We will now describe these steps in more detail.

\begin{figure}[]
\centering
\includegraphics[width=0.45\textwidth]{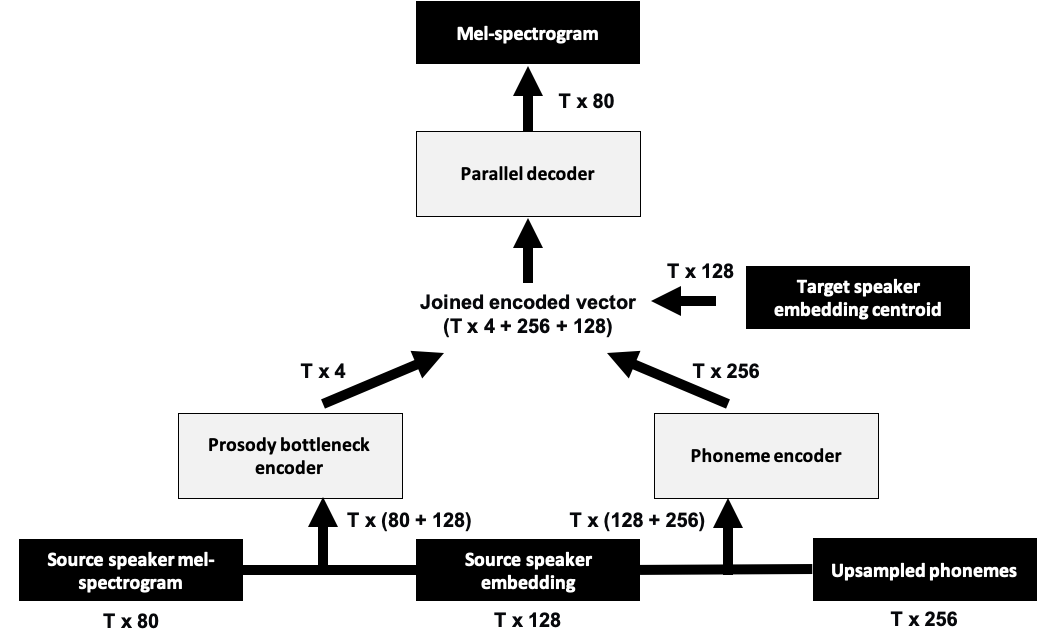}
\caption{VC model architecture}
\label{fig:copycat}
\end{figure}

\subsection{Data augmentation}
\label{ssec:dataaugmentation}

To create synthetic data for our experiments, we leverage the Copycat model \cite{karlapati2020copycat} as VC model (Fig. \ref{fig:copycat}). This model is a fine-grained prosody transfer model, aiming to maintain the prosody and linguistic content from a source audio while converting the speaker identity. It consists of three components: (1) a phoneme encoder that learns latent embeddings of the phonemes, (2) a prosody bottleneck encoder which disentangles the prosodic representation from the reference mel-spectrogram, and (3) a parallel decoder which generates a mel-spectrogram given the phoneme embeddings, the prosodic representation, and a provided speaker embedding. For more detailed information about the architecture and the functioning of the model, as well as for more information on the quality one can expect to obtain, we refer the reader to the Copycat paper \cite{karlapati2020copycat}.
 
We bring one modification to the model, by concatenating the speaker embeddings to the upsampled phonemes before passing the result to the phoneme encoder. This helps encode phonemes according to the speaker identity, and therefore aids the VC process. We find this change to reduce occurrences of speaker leakage.

The VC model is trained on one or more source speakers, plus the target speaker, making it possible to convert expressive samples of a style $X$ from a source speaker $S$ to a target speaker $T$. Despite the samples not being as perfect as recordings, they still are of great quality and helpful to train better TTS models, as shown in our evaluations.

\subsection{TTS model}
\label{ssec:ttsmodel}

\begin{figure}[]
\centering
\includegraphics[width=0.45\textwidth]{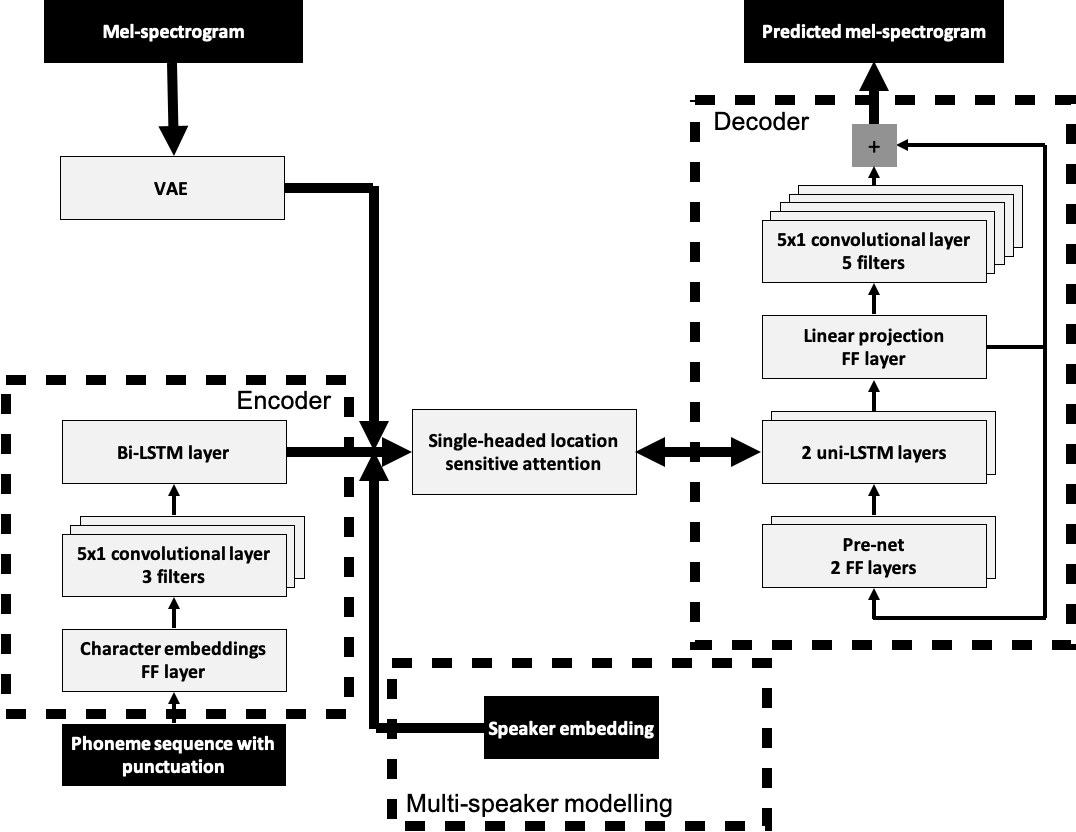}
\caption{TTS model architecture}
\label{fig:prosotron}
\vspace{-5mm}
\end{figure}

\subsubsection{Architecture}
\label{sssec:architecture}

Our TTS system is based on a Tacotron-like \cite{wang2017tacotron} structure with an additional variational auto-encoder (VAE) \cite{kingma2013auto} to capture prosody \cite{zhang2019learning, tyagi2019dynamic} (Fig. \ref{fig:prosotron}). Depending on the use case and to show the robustness of our approach across different scenarios, we either train a multi-style single-speaker or a single-style multi-speaker model. When training the latter model, we also add a pre-trained speaker embedding at the decoder level, which is obtained via the speaker classifier described in \cite{karlapati2020copycat}. The model is trained for 400k steps using a batch size of 32 with a Kullback-Leibler divergence loss for the VAE component, a L1-loss for the predicted mel-spectrograms and a cross-entropy stop-token loss. At inference time, the model is conditioned on the target speaker centroid VAE z-vector computed on the target style recordings. For generating raw audio waveforms from predicted mel-spectrograms, we use a parallel Wavenet universal neural vocoder \cite{oord2018parallel}.

\vspace*{-2mm}
\subsubsection{Fine-tuning}
\label{sssec:finetuning}

Once the model is trained on the entire dataset, we resume training the model in the exact same setting (i.e. same architecture, same losses, etc.), but only on the non-synthetic data for the target speaker in the target speaking style. This fine-tuning training is run for an additional 4k steps. Doing so, we are able to improve both the signal and segmental quality, as well as the expressiveness, without it being detrimental for stability/overfitting. The rationale behind this idea is that fine-tuning guides the model to focus on the space that matters the most. Unlike many existing low-resource TTS fine-tuning techniques \cite{tits2019exploring, chung2019semi, zhang2020unsupervised}, the target data is here already present in the so called pre-training step, making our fine-tuning step more of a refinement step.

\section{Experiments}
\label{sec:experiments}

\begin{table}[t]
\begin{center}
\scalebox{0.9}{
\begin{tabular}{|l|p{5.75cm}|}
\hline
\textbf{Notation} & \textbf{Description} \\ \hline
$S_{i,j,k}$                   & Dataset of real speaker $i$ in style $j$ consisting of $k$ hours                       \\
$S^{*}_{i,j,k}$                   & Dataset of synthetic/voice-converted speaker $i$ in style $j$ consisting of $k$ hours                      \\
VC($\lbrace{S_{i,j,k}}\rbrace$)                   & Voice conversion model trained on a set of datasets $S_{i,j,k}$                      \\
TTS$(\lbrace{S^{(*)}_{i,j,k}}\rbrace$)                   & TTS model trained on a set of datasets $S^{(*)}_{i,j,k}$                       \\
FT$(TTS, {S_{i,j,k}}$)                   & Fine-tuning model $TTS$ on dataset $S_{i,j,k}$                       \\ \hline
\end{tabular}}
\end{center}
\vspace{-5mm}
\caption{Notation used to describe experimental settings.}
\vspace{-5mm}
\label{tab:notation}
\end{table}

\subsection{Experimental settings}
\label{ssec:experimentalsettings}

We make use of professionally recorded American English internal datasets. To demonstrate the efficacy of the proposed methodology, we conducted experiments in two different data scenarios, for each of which we test on different expressive speaking styles (i.e. newscaster and conversational). For both scenarios, only 30 min of target recordings are used. We explain the settings on the basis of the notations of Table \ref{tab:notation}.

In the \textbf{single-speaker TTS - newscaster style} scenario, we have access to a large amount of target speaker data in a different speaking style (i.e. neutral). However, we have only one other different supporting speaker. We train next 2 pairs of VC \& multi-style single-speaker TTS models, one for target speaker \textit{ Female 1} and one for target speaker\textit{ Female 2}: \\

\hspace*{-5mm}1. VC$_{1}(S_{1,neutral,20h}, S_{1,news,0.5h}, S_{2,neutral,20h}, S_{2,news,7h}$) \\
$\hookrightarrow$ Convert $S_{2,news,7h}$ to $S^{*}_{1,news,7h}$  \\
$\hookrightarrow$ TTS$_{1}(S_{1,neutral,20h}, S_{1,news,0.5h}, S^{*}_{1,news,7h})$ \\
$\hookrightarrow$ FT$(TTS_{1}, S_{1,news,0.5h})$ \\
2. VC$_{2}(S_{1,neutral,20h}, S_{1,news,4h}, S_{2,neutral,20h}, S_{2,news,0.5h}$) \\
$\hookrightarrow$ Convert $S_{1,news,4h}$ to $S^{*}_{2,news,4h}$ \\
$\hookrightarrow$ TTS$_{2}(S_{2,neutral,20h}, S_{2,news,0.5h}, S^{*}_{2,news,4h})$ \\
$\hookrightarrow$ FT$(TTS_{2}, S_{2,news,0.5h})$ \\

In the \textbf{multi-speaker TTS - conversational style} scenario, we have a large amount of conversational data across various speakers (split evenly between male and female). However, we have no supporting data from the target speaker in other speaking styles. We train one single VC model, and one single-style multi-speaker TTS model for each of our 8 target speakers $S_{1-8}$, using 10 supporting speakers $S_{9-18}$: \\

\hspace*{-5mm}VC$(S_{1-8,conv,0.5h}, S_{9-15,conv,5h}, S_{16-18,conv,1.5h})$ \\
$\hookrightarrow$ $\forall i \in [1,8]$: Convert $S_{x,conv,5h}$ to $S^{*}_{i,conv,5h}$, with $x$ chosen from $S_{9-15,conv,5h}$ \\
$\hookrightarrow$ $\forall i \in [1,8]$: TTS$_{i}(S_{i,conv,0.5h}, S_{9-18,conv,\scriptstyle\sum}, S^{*}_{i,conv,5h})$ \\
$\hookrightarrow$ $\forall i \in [1,8]$: FT$(TTS_{i}, S_{i,conv,0.5h})$ \\

Experiments, omitted here due to a lack of space, found that converting data from more than one 5h-supporting speaker doesn't increase quality.

\subsection{Metrics and evaluation procedure}
\label{ssec:metrics}

Throughout the evaluations, we focus on 4 metrics to assess our models: (1) \textit{Signal quality} (How good is the signal quality of the samples?), (2) \textit{Style adequacy} (Do the samples match the desired style?), (3) \textit{Naturalness} (Do the samples sound natural?), and (4) \textit{Speaker similarity} (Do the samples sound like the target speaker?). For style adequacy and  speaker similarity, we provide the listeners a reference sample. We evaluate our models via MUSHRA tests \cite{itu20031534}. For each newscaster target speaker, 200 test samples have been held out and 200 testers have been recruited who evaluate each 10 test screens. For the conversational use case, 50 test utterances from each of the 8 target speakers are evaluated, with each test screen being rated 30 times. The evaluations are carried out using crowd-sourcing platforms. To check for statistical differences, we perform paired t-tests using the Holm-Bonferroni correction and a p-value of 0.05.

\vspace*{-2mm}
\subsection{Results}
\label{ssec:results}

\hspace*{5mm}\textit{Single-speaker TTS - Newscaster style}\\
\vspace*{-2mm}

\textbf{Component ablation} - In our first experiment, we show the (combined) improvements obtained by the two contributions of this paper: 1) using voice-converted data, and 2) fine-tuning the model on the target data. We train 4 types of models for the ablation study: (\textit{B}) a baseline TTS model, (\textit{B+FT}) the same model as (\textit{B}) but further fine-tuned on the newscaster recordings, (\textit{B+VC}) the same model as (\textit{B}) but including the voice-converted data in the training process, and (\textit{B+VC+FT}) the same model as (\textit{B+VC}) but further fine-tuned on the newscaster recordings.

Listener responses from the ablation study are shown in Table \ref{tab:table0}. These results show that the signal quality and the style adequacy of our methodology (\textit{B+VC+FT}) significantly outperforms all other models for both speakers, except for the \textit{Female 2} style adequacy, for which the improvement is not statistically different. Furthermore, we observe that the separate use of synthetic data and fine-tuning brings improvements in the majority of cases. However, the combination of both approaches brings the largest improvements. We hypothesise that including synthetic data helps the model to get a better picture of the type of data it has to produce, as it sees much more target data, despite it not being all real data. The fine-tuning then helps the model to focus on the actual target space more closely after having seen a lot of neutral and synthetic data.

\begin{table}[h!]
\begin{center}
\scalebox{0.85}{
\begin{tabular}{|l|l|l|l|l|}
\hline
\multirow{2}{*}{} & \multicolumn{2}{c|}{Female 1}                                    & \multicolumn{2}{c|}{Female 2}                                  \\ \cline{2-5} 
                  & \multicolumn{1}{c|}{Sig. Q.} & \multicolumn{1}{c|}{Style A.} & \multicolumn{1}{c|}{Sig. Q.} & \multicolumn{1}{c|}{Style A.} \\ \hline
Base (B)          & 69.9$\pm$1.1                     & 65.7$\pm$1.4                      & 67.0$\pm$1.0                     & 68.7$\pm$1.3                      \\
B+FT              & 69.7$\pm$1.0                     & 66.2$\pm$1.3                      & 69.3$\pm$1.0                     & \textbf{70.5}$\pm$1.3                      \\
B+VC              & 70.2$\pm$1.1                     & 66.6$\pm$1.4                      & 69.4$\pm$1.0                     & 70.0$\pm$1.3                      \\
\textbf{B+VC+FT}           & \textbf{\underline{72.9}}$\pm$1.0                     & \textbf{\underline{67.9}}$\pm$1.3                      & \textbf{\underline{71.2}}$\pm$1.0                     & 70.0$\pm$1.3          \\ \hline           
\end{tabular}}
\end{center}
\vspace{-5mm}
\caption{\footnotesize{Ablation study on usage of VC data and fine-tuning, showing average + 95\% CI MUSHRA scores for both signal quality and style adequacy in the single-speaker setting on 2 female speakers. An underlined value signifies a statistical difference between ($B$) and ($B+VC+FT$).}}
\label{tab:table0}
\end{table}

\textbf{Data ablation} - To understand how this methodology is affected by the amount of data from the target speaker in the target speaking style and compares to upper anchors, we performed further evaluations between different data reduced (\textit{DR}) and non-data reduced (\textit{non-DR}) scenarios. These evaluations compare 5 systems: (\textit{Recs}) recordings, (\textit{non-DR}) TTS models trained on neutral data and the full/non-data reduced newscaster dataset (i.e. 4h for \textit{Female 1}, 7h for \textit{Female 2}), (\textit{DR+VC+FT}) TTS models trained on neutral data, a reduced newscaster dataset (i.e. 45, 30 and 15 min), synthetic data generated via VC and fine-tuned for 4k steps, (\textit{DR}) TTS models trained on neutral data and a reduced newscaster dataset (i.e. 45, 30 and 15 min), and (\textit{Neutral}) a neutral TTS model which has been trained on the neutral dataset only.

It is expected that (\textit{DR}) performs worse than (\textit{non-DR}), as (\textit{DR}) is the same model as (\textit{non-DR}) but trained on less data (i.e. 5x to 28x less). We anticipate that (\textit{DR+VC+FT}), generated via our methodology, reduces (and potentially exceeds) the gap between (\textit{DR}) and (\textit{non-DR}). (\textit{DR+VC+FT}) observes the same amount of recorded data as (\textit{DR}), but makes additional use of the synthetic data and TTS model fine-tuning. (\textit{Recs}) is given as an upper anchor, while (\textit{Neutral}) is given as a lower anchor for the style adequacy metric. Every column in Table \ref{tab:table1} is made of one \textit{Female 1} and one \textit{Female 2} MUSHRA evaluation.

Table \ref{tab:table1} shows improvements for both speakers for the proposed model (\textit{DR+VC+FT}) over the baseline model (\textit{DR}), on both metrics and across 11 out of the 12 MUSHRA evaluations. Out of the 11, 8 are statistically significant improvements. For \textit{Female 1}, we notice that the gap between (\textit{DR}) and (\textit{non-DR}) is often much smaller than the equivalent gap for \textit{Female 2}. Hence, the improvements shown with our models (\textit{DR+VC+FT}) are sometimes able to not only improve over (\textit{DR}) but also (\textit{non-DR}). For \textit{Female 2}, (\textit{DR+VC+FT}) never reaches the quality of (\textit{non-DR}), but the gap between (\textit{DR}) and (\textit{non-DR}) almost always gets vastly reduced. While our methodology works remarkably well for all data scenarios, for the (lowest-resource) 15-minute \textit{Female 2} scenario, it is even the reason why a newscaster model can be built in the first place. Without the use of synthetic data and fine-tuning, the model is only able to generate neutral speech, as can be seen by the style adequacy score that is similar to the one of the neutral TTS model.

\begin{table}[h!]
\begin{center}
\scalebox{0.83}{
\begin{tabular}{|l|
>{\columncolor[HTML]{EFEFEF}}l |c|
>{\columncolor[HTML]{EFEFEF}}l |c|
>{\columncolor[HTML]{EFEFEF}}l |c|}
\hline
\cellcolor[HTML]{9B9B9B} & \multicolumn{3}{c|}{\cellcolor[HTML]{9B9B9B}Signal quality} & \multicolumn{3}{c|}{\cellcolor[HTML]{9B9B9B}Style adequacy} \\ \hline
\#min news recs          & 45                 & 30                 & 15                & 45                 & 30                 & 15                \\ \hline
\multicolumn{7}{|c|}{\cellcolor[HTML]{C0C0C0}Female 1}                                                                                                   \\ \hline
Recs               & 76.9               & 77.8               & 78.0              & 74.7               & 78.9               & 78.1              \\
non-DR (4h)              & 71.4               & 71.7               & 71.9              & 69.7               & 72.7               & 68.8              \\
\textbf{DR+VC+FT}             & \textbf{\underline{71.9}}               & \textbf{\underline{73.0} }              & \textbf{69.0}              & \textbf{68.3}               & \textbf{\underline{72.6}}               & 63.9              \\
DR                       & 69.4               & 69.2               & 68.9              & 67.3               & 69.0               & \textbf{64.0}              \\
Neutral                  & X                  & X                  & X                 & 61.8               & 63.9               & 60.7              \\ \hline
\multicolumn{7}{|c|}{\cellcolor[HTML]{C0C0C0}Female 2}                                                                                                 \\ \hline
Recs               & 75.0               & 74.0               & 71.7              & 79.2               & 77.2               & 76.7              \\
non-DR (7h)              & 71.0               & 70.2               & 69.2              & 72.4               & 71.0               & 71.4              \\
\textbf{DR+VC+FT}             & \textbf{\underline{68.6}}               & \textbf{\underline{66.8}}               & \textbf{\underline{67.5}}              & \textbf{\underline{69.2}}               & \textbf{66.7}               & \textbf{\underline{67.7}}              \\
DR                       & 58.8               & 62.7               & 66.5              & 66.9               & 66.4               & 58.1              \\
Neutral                  & X                  & X                  & X                 & 61.5               & 61.4               & 56.0             \\ \hline
\end{tabular}}
\end{center}
\vspace{-5mm}
\caption{\footnotesize{Average MUSHRA scores for signal quality and style adequacy in the single-speaker setting, showing the performance of our methodology ($DR+VC+FT$) in different data reduced scenarios and for 2 female speakers. An underlined value signifies a statistical difference between ($DR$) and ($DR+VC+FT$).}}
\label{tab:table1}
\end{table}

\newpage
\textit{Multi-speaker TTS - Conversational style} \\
\vspace*{-2mm}

Finally, we demonstrate our methodology in a different scenario, to illustrate its robustness. The target speakers have different amounts of recordings available, enabling us to understand the comparative effects of data reduction (i.e. how does reducing a voice to 30 min of speech data compared to having 45 min, 1.5h or 5h of recordings available?). We compare here two multi-speaker models: (\textit{non-DR}) trained using all data available for all supporting and target speakers, and (\textit{DR+VC+FT}) trained for each target speaker using only 30 min of their recordings plus all data from the supporting speakers with the proposed approach.

Table \ref{tab:table2} shows that given just 30 min of speech data the proposed methodology is found to be significantly better than voices trained on the full amount of data for the target speaker. This level of quality is achieved for all 8 speakers evaluated, and this even when 90\% of reduction in data was applied (i.e. from 5h of data to 30 min). In addition, we observe that this is achieved without degrading perceived speaker similarity. We believe results are even better for this single-style multi-speaker setting, as synthetic data becomes increasingly valuable in the absence of supporting neutral target speaker data.

\begin{table}[h!]
\begin{center}
\scalebox{0.85}{
\begin{tabular}{|l|l|c|c|}
\hline
\rowcolor[HTML]{9B9B9B} 
\multicolumn{2}{|c|}{}         & Naturalness      & Speaker similarity      \\ \hline
\rowcolor[HTML]{C0C0C0} 
\multicolumn{4}{|c|}{\cellcolor[HTML]{C0C0C0}SMALL - Reducing 45 min to 30 min} \\ \hline
                                         & Recs            & 78.80$\pm$0.73            & 68.10$\pm$0.65          \\ \cline{2-4} 
                                         & non-DR          & 51.78$\pm$0.76            & 65.61$\pm$0.67         \\ \cline{2-4} 
\multirow{-3}{*}{4 speakers}      & \textbf{DR+VC+FT}              & \textbf{\underline{56.86}}$\pm$0.72            & \textbf{\underline{66.80}}$\pm$0.66          \\ \hline
\rowcolor[HTML]{C0C0C0} 
\multicolumn{4}{|c|}{\cellcolor[HTML]{C0C0C0}MEDIUM - Reducing 1.5h to 30 min} \\ \hline
                                         & Recs            & 79.86$\pm$1.02             & 69.78$\pm$0.89         \\ \cline{2-4} 
                                         & non-DR          & 56.58$\pm$1.04            & \textbf{67.45}$\pm$0.91         \\ \cline{2-4} 
\multirow{-3}{*}{2 speakers}      & \textbf{DR+VC+FT}              & \textbf{\underline{58.72}}$\pm$1.02            & 67.41$\pm$0.90         \\ \hline
\rowcolor[HTML]{C0C0C0} 
\multicolumn{4}{|c|}{\cellcolor[HTML]{C0C0C0}LARGE - Reducing 5h to 30 min}         \\ \hline
                                         & Recs      & 81.65$\pm$1.00            & 67.83$\pm$0.94         \\ \cline{2-4} 
                                         & non-DR          & 55.16$\pm$1.06            & 66.25$\pm$0.95         \\ \cline{2-4} 
\multirow{-3}{*}{2 speakers}      & \textbf{DR+VC+FT}              & \textbf{\underline{59.39}}$\pm$0.97            & \textbf{66.72}$\pm$0.93         \\ \hline
\end{tabular}}
\end{center}
\vspace{-5mm}
\caption{\footnotesize{Average + 95\% CI MUSHRA scores for naturalness and speaker similarity in the multi-speaker setting, showing the performance of our methodology ($DR+VC+FT$) in different data reduced scenarios and for 8 speakers (evenly split between male and female). An underlined value signifies a statistical difference between ($non-DR$) and ($DR+VC+FT$).}}
\label{tab:table2}
\vspace{-5mm}
\end{table}

\vspace*{-1mm}
\section{Conclusion}
\label{sec:conclusion}
\vspace*{-1mm}

We propose a novel methodology to build high-quality expressive TTS models, while having only little expressive recordings for the target speaker. The methodology consists of: 1) generating synthetic data via a VC model, 2) incorporating this converted data in training a Tacotron-like TTS model, and 3) fine-tuning this TTS model on the target data. The proposed approach brings improvements in signal quality, naturalness and style adequacy, whilst suffering no drop in speaker similarity. Whilst many VC papers exist, as far as we are aware, there has been no work on the use of VC-created data to train TTS models. The successful use of such data to train TTS models means a reduced cost of recording data, as well as a faster and easier expansion of TTS voices.

\pagebreak

\bibliographystyle{IEEEbib}
\small{\bibliography{strings}}

\begin{thebibliography}{10}

\bibitem{oord2016wavenet}
Aaron van~den Oord et~al.,
\newblock ``Wavenet: A generative model for raw audio,''
\newblock {\em arXiv preprint arXiv:1609.03499}, 2016.

\bibitem{wang2017tacotron}
Yuxuan Wang et~al.,
\newblock ``Tacotron: Towards end-to-end speech synthesis,''
\newblock {\em arXiv preprint arXiv:1703.10135}, 2017.

\bibitem{sotelo2017char2wav}
Jose Sotelo et~al.,
\newblock ``Char2wav: End-to-end speech synthesis,''
\newblock 2017.

\bibitem{skerry2018towards}
RJ~Skerry-Ryan et~al.,
\newblock ``Towards end-to-end prosody transfer for expressive speech synthesis
  with tacotron,''
\newblock {\em arXiv preprint arXiv:1803.09047}, 2018.

\bibitem{kalchbrenner2018efficient}
Nal Kalchbrenner et~al.,
\newblock ``Efficient neural audio synthesis,''
\newblock {\em arXiv preprint arXiv:1802.08435}, 2018.

\bibitem{oord2018parallel}
Aaron Oord et~al.,
\newblock ``Parallel wavenet: Fast high-fidelity speech synthesis,''
\newblock in {\em International conference on machine learning}. PMLR, 2018,
  pp. 3918--3926.

\bibitem{chung2019semi}
Yu-An Chung et~al.,
\newblock ``Semi-supervised training for improving data efficiency in
  end-to-end speech synthesis,''
\newblock in {\em ICASSP 2019-2019 IEEE International Conference on Acoustics,
  Speech and Signal Processing (ICASSP)}. IEEE, 2019, pp. 6940--6944.

\bibitem{gibiansky2017deep}
Andrew Gibiansky et~al.,
\newblock ``Deep voice 2: Multi-speaker neural text-to-speech,''
\newblock in {\em Advances in neural information processing systems}, 2017, pp.
  2962--2970.

\bibitem{jia2018transfer}
Ye~Jia et~al.,
\newblock ``Transfer learning from speaker verification to multispeaker
  text-to-speech synthesis,''
\newblock in {\em Advances in neural information processing systems}, 2018, pp.
  4480--4490.

\bibitem{tits2019exploring}
No{\'e} Tits et~al.,
\newblock ``Exploring transfer learning for low resource emotional tts,''
\newblock in {\em Proceedings of SAI Intelligent Systems Conference}. Springer,
  2019, pp. 52--60.

\bibitem{latorre2019effect}
Javier Latorre et~al.,
\newblock ``Effect of data reduction on sequence-to-sequence neural tts,''
\newblock in {\em ICASSP 2019-2019 IEEE International Conference on Acoustics,
  Speech and Signal Processing (ICASSP)}. IEEE, 2019, pp. 7075--7079.

\bibitem{chen2019end}
Yuan-Jui Chen et~al.,
\newblock ``End-to-end text-to-speech for low-resource languages by
  cross-lingual transfer learning.,''
\newblock in {\em Interspeech}, 2019, pp. 2075--2079.

\bibitem{zhang2020unsupervised}
Haitong Zhang et~al.,
\newblock ``Unsupervised learning for sequence-to-sequence text-to-speech for
  low-resource languages,''
\newblock {\em arXiv preprint arXiv:2008.04549}, 2020.

\bibitem{perez2017effectiveness}
Luis Perez et~al.,
\newblock ``The effectiveness of data augmentation in image classification
  using deep learning,''
\newblock {\em arXiv preprint arXiv:1712.04621}, 2017.

\bibitem{mikolajczyk2018data}
Agnieszka Miko{\l}ajczyk et~al.,
\newblock ``Data augmentation for improving deep learning in image
  classification problem,''
\newblock in {\em 2018 international interdisciplinary PhD workshop (IIPhDW)}.
  IEEE, 2018, pp. 117--122.

\bibitem{shorten2019survey}
Connor Shorten et~al.,
\newblock ``A survey on image data augmentation for deep learning,''
\newblock vol. 6, no. 1, pp. 60, 2019.

\bibitem{ko2015audio}
Tom Ko et~al.,
\newblock ``Audio augmentation for speech recognition,''
\newblock in {\em Sixteenth Annual Conference of the International Speech
  Communication Association}, 2015.

\bibitem{ko2017study}
Tom Ko et~al.,
\newblock ``A study on data augmentation of reverberant speech for robust
  speech recognition,''
\newblock in {\em 2017 IEEE International Conference on Acoustics, Speech and
  Signal Processing (ICASSP)}. IEEE, 2017, pp. 5220--5224.

\bibitem{park2019specaugment}
Daniel~S Park et~al.,
\newblock ``Specaugment: A simple data augmentation method for automatic speech
  recognition,''
\newblock {\em arXiv preprint arXiv:1904.08779}, 2019.

\bibitem{xu2020lrspeech}
Jin Xu et~al.,
\newblock ``Lrspeech: Extremely low-resource speech synthesis and
  recognition,''
\newblock in {\em Proceedings of the 26th ACM SIGKDD International Conference
  on Knowledge Discovery \& Data Mining}, 2020, pp. 2802--2812.

\bibitem{mohammadi2017overview}
Seyed~Hamidreza Mohammadi et~al.,
\newblock ``An overview of voice conversion systems,''
\newblock {\em Speech Communication}, vol. 88, pp. 65--82, 2017.

\bibitem{hsu2017voice}
Chin-Cheng Hsu et~al.,
\newblock ``Voice conversion from unaligned corpora using variational
  autoencoding wasserstein generative adversarial networks,''
\newblock {\em arXiv preprint arXiv:1704.00849}, 2017.

\bibitem{lorenzo2018voice}
Jaime Lorenzo-Trueba et~al.,
\newblock ``The voice conversion challenge 2018: Promoting development of
  parallel and nonparallel methods,''
\newblock {\em arXiv preprint arXiv:1804.04262}, 2018.

\bibitem{kaneko2019cyclegan}
Takuhiro Kaneko et~al.,
\newblock ``Cyclegan-vc2: Improved cyclegan-based non-parallel voice
  conversion,''
\newblock in {\em ICASSP 2019-2019 IEEE International Conference on Acoustics,
  Speech and Signal Processing (ICASSP)}. IEEE, 2019, pp. 6820--6824.

\bibitem{karlapati2020copycat}
Sri Karlapati et~al.,
\newblock ``Copycat: Many-to-many fine-grained prosody transfer for neural
  text-to-speech,''
\newblock {\em arXiv preprint arXiv:2004.14617}, 2020.

\bibitem{kingma2013auto}
Diederik~P Kingma et~al.,
\newblock ``Auto-encoding variational bayes,''
\newblock {\em arXiv preprint arXiv:1312.6114}, 2013.

\bibitem{zhang2019learning}
Ya-Jie Zhang et~al.,
\newblock ``Learning latent representations for style control and transfer in
  end-to-end speech synthesis,''
\newblock in {\em ICASSP 2019-2019 IEEE International Conference on Acoustics,
  Speech and Signal Processing (ICASSP)}. IEEE, 2019, pp. 6945--6949.

\bibitem{tyagi2019dynamic}
Shubhi Tyagi et~al.,
\newblock ``Dynamic prosody generation for speech synthesis using
  linguistics-driven acoustic embedding selection,''
\newblock {\em arXiv preprint arXiv:1912.00955}, 2019.

\bibitem{itu20031534}
RB~ITU-R,
\newblock ``1534-1, method for the subjective assessment of intermediate
  quality levels of coding systems (mushra),”,''
\newblock {\em International Telecommunication Union}, 2003.

\end{thebibliography}

\end{document}